\documentclass{ijcss}

\usepackage{amsmath,amsthm,color,bm,cite}
\usepackage{amssymb}
\usepackage{graphicx}
\usepackage{ifthen}

\begin{document}

\bibliographystyle{acm}

%
\newcommand{\Eqref}[1]{(\ref{#1})}
\newcommand{\kc}{k_{\textrm{\tiny C}}} \newcommand{\Nc}{N_{\textrm{\tiny C}}}
\newcommand{\kin}{k_{\textrm{\scriptsize in}}}
\newcommand{\pin}{p_{\textrm{\scriptsize in}}}
\newcommand{\pout}{p_{\textrm{\scriptsize out}}}
\newcommand{\kout}{k_{\textrm{\scriptsize out}}}
\newcommand{\nD}{n_{D}}
\newcommand{\ND}{N_{D}}
\newcommand{\rank}{\textrm{rank}}
%

\markboth {Controllability of a swarm with topologically interacting
  autonomous agents} 
{M. Komareji \& R. Bouffanais} 


\title{Controllability of a swarm of topologically interacting autonomous
  agents}



\author{Mohammad Komareji}{1} \author[bouffanais@sutd.edu.sg]{Roland
  Bouffanais}{1}

\affiliation{1}{Singapore University of Technology and Design}{20 Dover Drive,
  Singapore 138682}


\begin{abstract}
  Controllability of complex networks has been the focal point of many recent
  studies in the field of complexity~\cite{ref:ctrl,yuan13:_exact}. These
  landmark advances shed a new light on the dynamics of natural and
  technological complex systems~\cite{ref:ctrl,yuan13:_exact}. Here, we
  analyze the controllability of a swarm of autonomous self-propelled agents
  having a topological neighborhood of interactions, applying the analytical
  tools developed for the study of the controllability of arbitrary complex
  directed networks. To this aim we thoroughly investigate the structural
  properties of the swarm signaling network which is the information transfer
  channel underpinning the dynamics of agents in the physical space. Our
  results show that with 6 or 7 topological neighbors, every agent not only
  affects, but is also affected by all other agents within the
  group~\cite{cavagna:scale}. More importantly, still with 6 or 7 topological
  neighbors, each agent is capable of full control over all other agents. This
  finding is yet another argument justifying the particular value of the
  number of topological neighbors observed in field observations with flocks
  of starlings~\cite{cavagna}.
  \msccodes 82C22, 94C15, 05C40

  \corrauth \received Date of submission \published Date of publication \doi
  DOI number
\end{abstract}

\newpage

\section{Introduction}

%
The connectedness of the swarm signaling network (SSN), the swarm's
information transfer channel, has been shown to be a sufficient condition for
an agent within the swarm to affect and get affected by some if not all agents
of the group~\cite{komareji13:_resil_contr_dynam_collec_behav}. However, in
many occasions, one or more informed agents need to be able to drive the swarm
to a certain global state, and usually within finite time. This is better
understood when considering two biological systems such as a flock of birds or
a school of fish. For instance, evasive maneuvers triggered by a predator
approaching or by collision avoidance collective responses are induced by one
or a few agents perceiving the threat and responding to it. Those agents are
said to be informed since they involuntarily have a privileged
access to out-of-the-swarm informational signals. Moreover, these few agents
effectively are driver agents: they are able to control the entire swarm by
bringing the other agents to swiftly respond to a threat that they are not
directly detecting. It is worth adding that those driver agents do not possess
any ``super'' power of any sort but they simply temporarily become informed
``leaders'' as they happened to have discerned the danger first; any other
agent in the swarm could be driving the group as long as it is subjected to
specific external cues which are not made available globally to the whole
swarm. Therefore, controllability is a vital factor for a swarm to robustly
and effectively perform a dynamic collective response benefiting the majority
of the group members. In this paper, we analyze the controllability of a
dynamic swarm by tapping into network-theoretic concepts to represent the
dynamic complex network of interactions underlying the dynamics of the
collective in the case of topological interactions.

\section{The Swarming Model}
\label{model}

%
The model we investigate here, as a simple representation of swarming---is
composed of self-propelling agents moving about a two-dimensional plane with
constant speed, $v_0$, similarly to the Vicsek's
model~\cite{vicsek95:_novel}. However, the neighborhood of interactions is not
metric but instead is topological~\cite{cavagna}. The topological character of
the neighborhood of interactions has a tremendous impact on the properties of
interagent connectivity, in particular with the induced asymmetry in the
relationship whereby if agent $j$ is in the neighborhood of agent $i$, then
$i$ is not necessarily in the neighborhood of $j$, i.e. the interaction is
directed.

For simplicity, we assume that each agent $i$ is fully characterized by one
unique state variable $\theta_i$, its velocity $\mathbf{v}_{i} = v_0
\cos\theta_i \hat{x} + v_0 \sin\theta_i \hat{y}$, or equivalently its velocity
direction $\theta_i$, the speed $v_0$ being constant. The local
synchronization protocol---based on relative states that prevents any
singularity such as those reported with the original Vicsek's
model~\cite{wei08:_singul} from occuring---is strictly equivalent to a local
alignment rule, which mathematically can be stated as:
\begin{equation}\label{eq:consensus}
  \dot{\theta_i}(t) = \frac{1}{\left| \mathcal{N}_i(t) \right|}\sum\limits_{j \in
    \mathcal{N}_i(t)} w_{ij}(\theta_j(t) - \theta_i(t)),
\end{equation}
where $\mathcal{N}_i(t)$ is the time-dependent set of outdegree neighbors in
the agent $i$'s topological neighborhood of interaction, with cardinal number
$\left| \mathcal{N}_i(t) \right|$, and $w_{ij}$ is the binary weight of the
$i-j$ communication link. Note that in some models, $w_{ij}$ can take a more
complicated form than our binary choice~\cite{mirabet,cucker,bode2}. Using the
$k$-nearest neighbor rule for the topological neighborhood of interactions, we
have $\left| \mathcal{N}_i(t) \right|=k$ and the following dynamical equation
for each individual agent in the swarm:
\begin{equation}\label{smotion}
  \dot{\theta}_{i} = \frac{1}{k}\left[(\theta_j - \theta_{i}) + \cdots +
    (\theta_{j+k-1}  - \theta_{i})\right],
\end{equation}
where $\theta_j,\cdots,\theta_{j+k-1}$ are its $k$-nearest neighbors' velocity
directions.

\section{The Swarm Controllability}
\label{control}

\begin{figure}[htbp]
  \centering\label{fign}
  \includegraphics[width=0.38\textwidth]{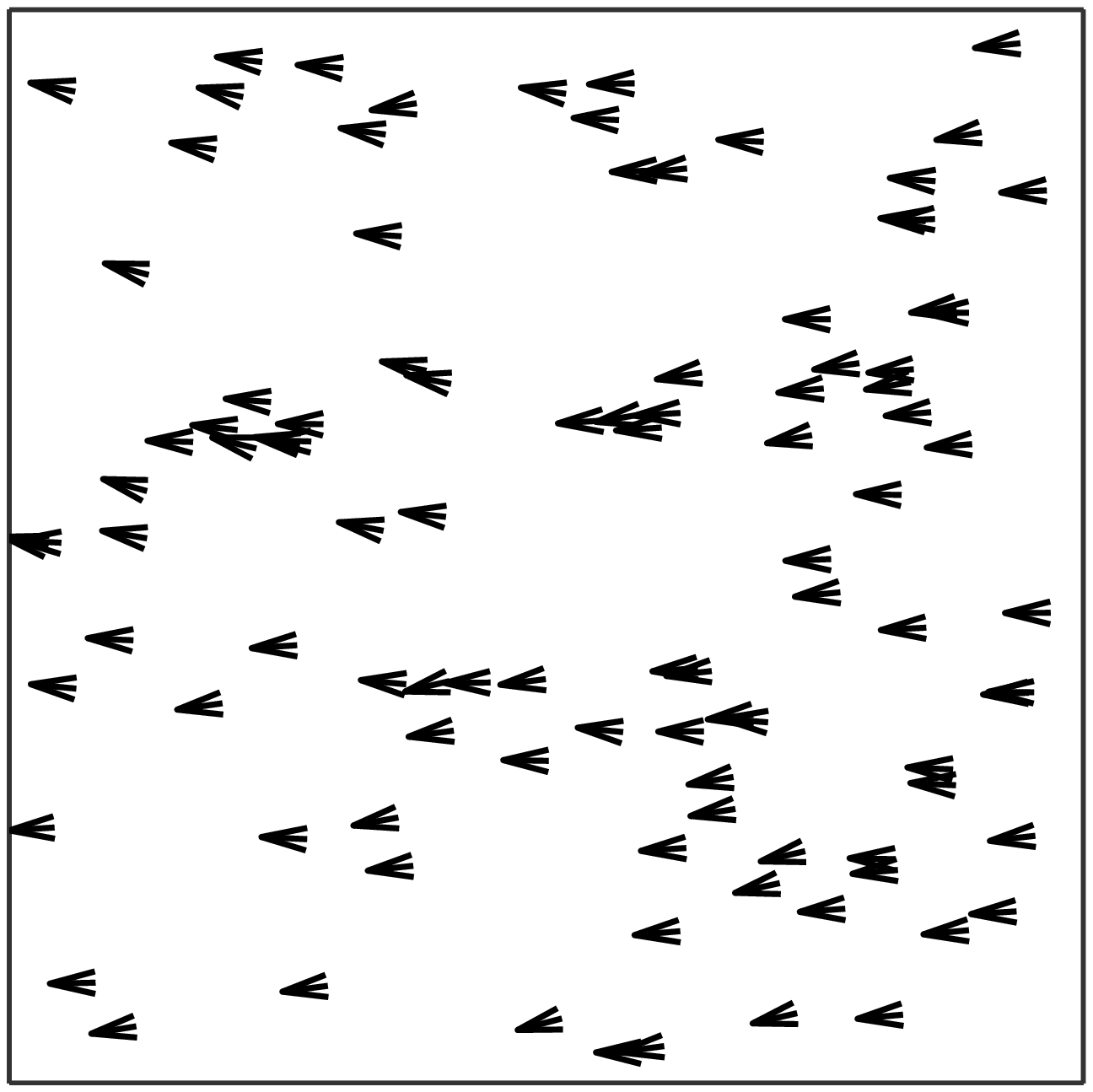}\\
  \includegraphics[width=0.38\textwidth]{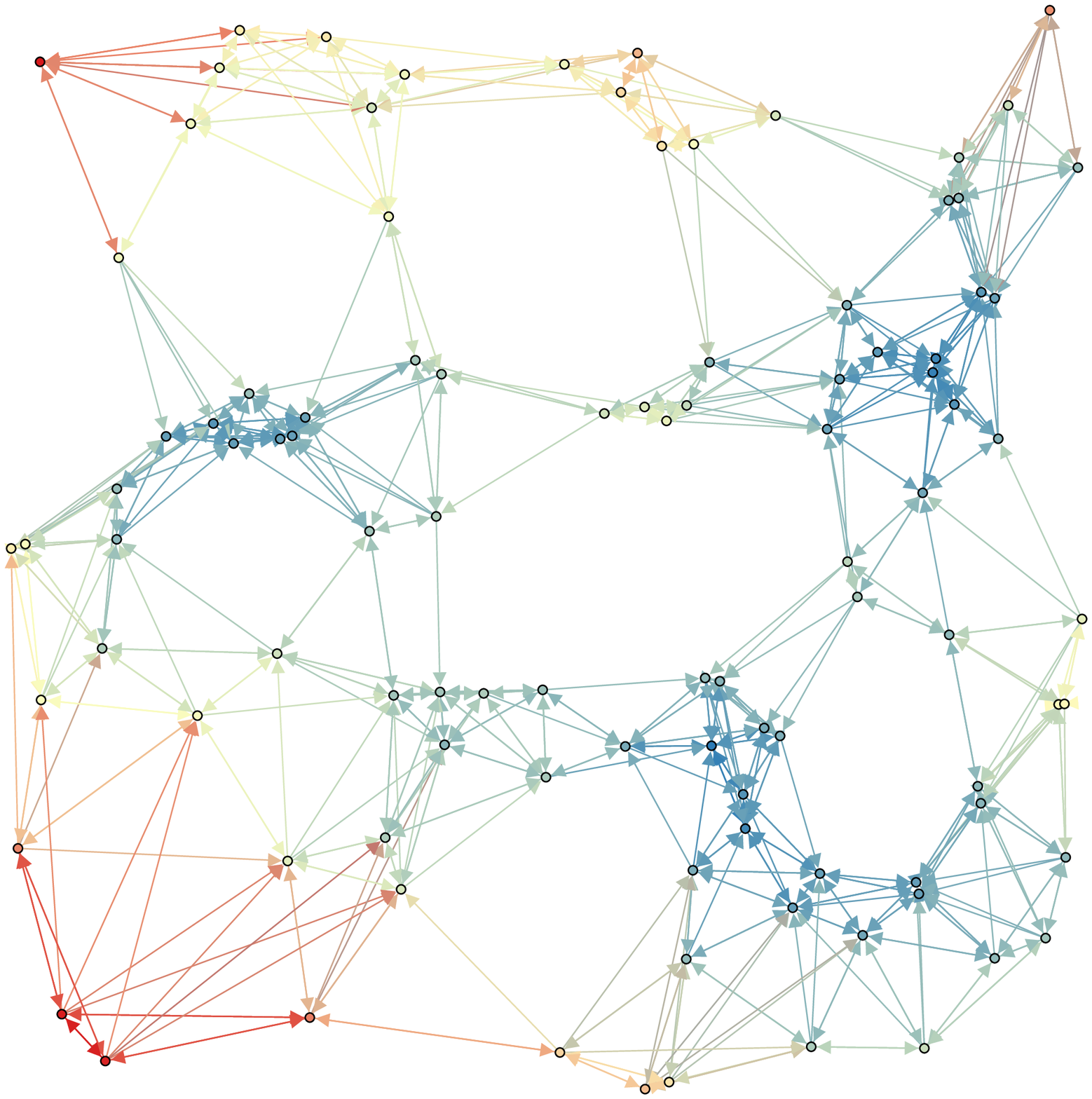}\\
  \includegraphics[width=0.38\textwidth]{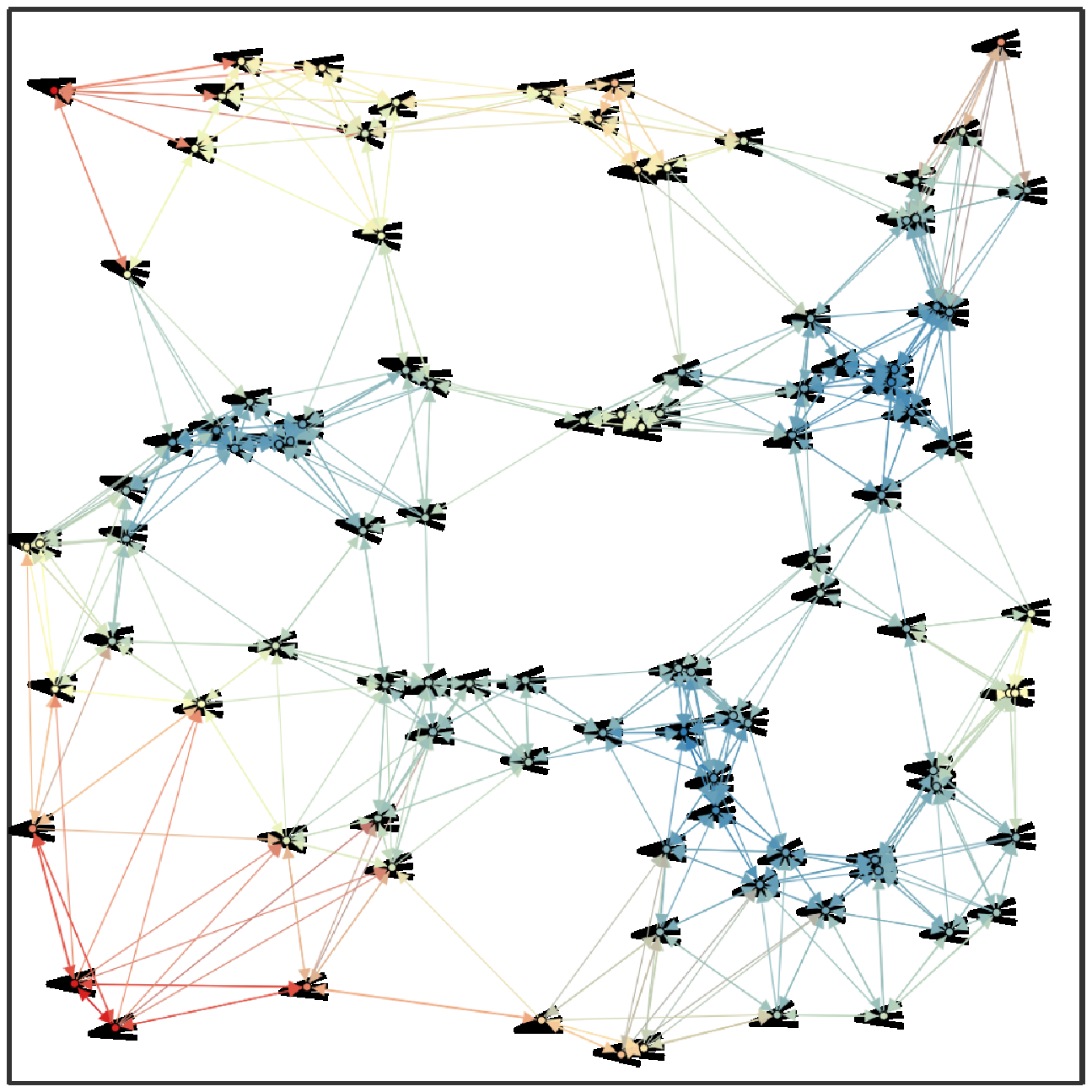}
  \caption{(Top) Physical view: snapshot of a swarm of $N=100$ topologically-interacting
    individuals traveling at constant speed ($v_0=0.03$) in a 2D square domain
    ($10\times10$) with periodic boundaries; each agent interacts
    topologically with $k=7$ neighbors. (Middle) Network view: the
    associated swarm signaling network (SSN); the nodes and edges are colored
    according to the topological distance (increasing topological distance from
    blue to red). (Bottom) Combined view: the swarm overlaid with the SSN.}
\end{figure}

%
In order to analyze the swarm controllability, we need to identify the SSN
which is the information transfer channel in the swarm underlying the dynamics
of the interacting agents. The dynamics of the agents in the two-dimensional
physical space is intricately coupled to the dynamics of the SSN. It is easy
to verify that the SSN is a switching $k$-nearest neighbor
digraph~\cite{komareji13:_resil_contr_dynam_collec_behav,ref:eppstein,ref:balister1,ref:balister2}
as agents are forced to interact with their $k$-nearest neighbors within the
evolving swarm. Consequently, the global swarm dynamical model can be recast
as
\begin{equation}\label{eq:main}
  \dot{\mathbf{\Theta}}(t) = \frac{1}{k} (-L) \mathbf{\Theta}(t),
\end{equation}
where $\mathbf{\Theta}(t)=\left[\theta_1(t), \cdots , \theta_N(t)
\right]^\textrm{T}$ is the vector of velocity directions of all agents and $L$
is the matrix of the graph Laplacian associated with the SSN based on the
outdegree. Note that given the $k$-nearest neighbor rule used for the
topological neighborhood of interactions, the outdegree for every single node
is constant and equal to $k$. Figure~\ref{fign}(top) depicts a snapshot of the
collective migration of a swarm
comprising $N=100$ topologically-interacting agents moving in a
two-dimensional square domain subjected to periodic boundary
conditions. The associated signaling network is shown in Fig.~\ref{fign}(middle)
with the nodes representing the traveling agents at their exact location in the physical
space, while the edges represent the directed topological interactions
between individuals.

Recently, the field of complex networks have seen the emergence of new general
theories and tools related to the controllability of such networks. The two
most prominent controllability tools are: (i) the structural controllability
framework developed by Liu~\textit{et
  al.}~\cite{ref:ctrl}, and (ii) the exact controllability framework very
recently introduced by Yuan~\textit{et al.}~\cite{yuan13:_exact}. In applying
both the exact and the structural controllability tools, one has access to the
details of the extent of the swarm's controllability, however, it requires
adapting the dynamical governing equations for the swarm to these
frameworks. This is the aim of the following lemma.

\smallskip
\noindent\textbf{Lemma.}\quad \itshape The controllability of the system
governed by~Eq.~\eqref{eq:main} is equivalent to the controllability of the
system
\begin{equation}\label{eq:ctrl}
  \dot{\mathbf{\Theta}}(t) = A \mathbf{\Theta}(t),
\end{equation}
where $A$ is the adjacency matrix of the SSN, whose graph Laplacian matrix $L$
appears in Eq.~\eqref{eq:main}.  \normalfont

\begin{proof}
  The number $N_D$ of driver nodes---a.k.a. unmatched nodes---in the system
  governed by Eq.~\eqref{eq:main} is determined as follows:
  \begin{equation}
    N_D = \max_i \left\{N - \rank\left(\delta_i I + \frac{1}{k} L\right)\right\}, 
  \end{equation}
  where $\delta_i$ is the $i$-th eigenvalue of $\tilde{L}=-L/k$.

  Given the definition of the matrix of the graph Laplacian and the
  topological nature of the inter-agent interactions, one can write:
  \begin{equation}\label{eq:equal}
    \delta_i I + \frac{1}{k} L = \delta_i I + \frac{1}{k} ( D - A) = \delta_i I + \frac{1}{k} (k I - A) = (\delta_i + 1) I - \frac{1}{k} A.
  \end{equation}
  It is easy to check that both transition matrices in Eqs.~\eqref{eq:main}
  and~\eqref{eq:ctrl} share the same eigenvectors. Thus, their corresponding
  eigenvalues are associated as
  \begin{equation}\label{eq:eig}
    \delta_i = \frac{1}{k} \lambda_i - 1,
  \end{equation}
  where $\lambda_i$ is the $i$-th eigenvalue of the adjacency matrix $A$ of
  the SSN.  Given Eqs.~\eqref{eq:equal} and \eqref{eq:eig}, we can conclude
  that
  \begin{equation}
    \rank\left(\delta_i I + \frac{1}{k} L\right) = \rank(\lambda_i I - A).
  \end{equation}
\end{proof}

\begin{figure}[htbp]
  \centering\label{figec}
  \includegraphics[width=0.7\textwidth]{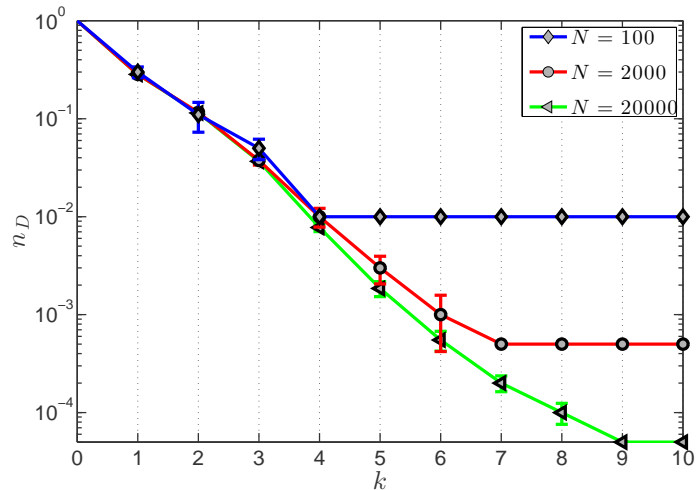}
  \caption{Density of required driver agents for a swarm with
    topologically-interacting members vs. the number of neighbors ($k$) for
    three different swarm populations ($N$). Results
    applying the exact controllability tool were collected for 10 distinct SSNs at each
    data point. The average density of driver nodes is calculated and the
    related standard deviations are illustrated by means of errorbars.}
\end{figure}

In our swarming model, interactions among all individual agents are either
``on'' or ``off'' depending on whether the pair of agents are topologically
interacting or not---or equivalently we can say that the weights of the
constituent links are binary numbers, 0 or 1. This highlights the fact that
link weights are not free independent parameters in our SSN model. Hence, the
exact controllability framework looks suitable to be applied to our
problem~\cite{yuan13:_exact}. Figure~\ref{figec} shows the results of the
exact controllability analysis of the dynamical swarm at any given point in
time. One can see that the number of driver nodes decrease exponentially as
$k$, the number of agents in the topological neighborhood, increases. We can
conclude through these results that if the number of nearest neighbors reaches
a value around $6$ to $8$---typical values for the number of topological
neighbors observed by Ballerini \textit{et al.}~\cite{cavagna} during field
experiments with bird flocks---every agent not only affects and is affected by
all other agents within the group, but more importantly, is capable of full
control over all other agents, i.e. the swarm.

One concern that should be addressed regarding the above results on the number
of driver nodes and the overall controllability of the swarm is associated
with the dynamic nature of the SSN. Since the SSN is intrinsically a switching
network---at each instant a certain number of links are broken while the exact
same number of edges are created due to the motion of the agents in the
physical space---one can prove that it is controllable at each instant,
assuming of course a high-enough value for $k$, for example around $6$ to
$8$. If that is the case, it is known from control theory associated with
dynamic hybrid systems that the overall switching dynamical system is
controllable~\cite{switchctrl,switchctrl2}. However, if the value of $k$ is
not large enough to have a controllable swarm at each instant, then this
analysis reveals a lower bound for the control centrality of each single
agent, i.e. the ability of a single agent to control the whole
swarm~\cite{ctrlcentral}.

In either natural or artificial swarms it is more realistic to have non-binary
weights for communication links in order to model the imperfection of the
information transfer channel. Thus, it is necessary to consider how the swarm
controllability is affected by changing the weights of edges of the
SSN. Moreover, such an study would reveal the efficiency of our simple model
in analyzing the swarm controllability associated with realistic cases. To
that end, we further perform a structural controllability analysis of the
swarm.

A system's structural controllability is to a great extent encoded in the
underlying degree distribution, $p(\kin,\kout)$. That is, the number of driver
agents is determined mainly by the number of incoming and outgoing links each
node of the SSN has, and is independent of where those links point
at~\cite{ref:ctrl}. As mentioned before and by construction, the outdegree
distribution of the SSN is a Dirac delta distribution, while its indegree
distribution very much resembles the one of a $k$-nearest random
digraph~\cite{komareji13:_resil_contr_dynam_collec_behav}, namely a Poisson
distribution associated with mean degree $k$. To allow for an analytical study
of the structural controllability of the swarm, we therefore consider the
following degree distributions:
\begin{equation}\label{eq:dd}
  \begin{tabular}{c}
    $\pout(\kout)=\delta(\kout-k),$\\
    $\pin(\kin)= \frac{k^{\kin}}{\kin!} e^{-k}.$
  \end{tabular}
\end{equation}
Given the above discussion, the following lemma provides a key and
useful result originating from the structural controllability
framework~\cite{komareji13:_resil_contr_dynam_collec_behav}.

\smallskip
\noindent\textbf{Lemma.}\quad \itshape The number of driver agents of the
system governed by Eq.~\eqref{eq:ctrl} at each time instant is given by $\ND
\approx \frac{N}{2} e^{-k}$, in the large $k$ limit.\normalfont
\begin{figure}[htbp]
  \centering\label{figsc}
  \includegraphics[width=0.7\textwidth]{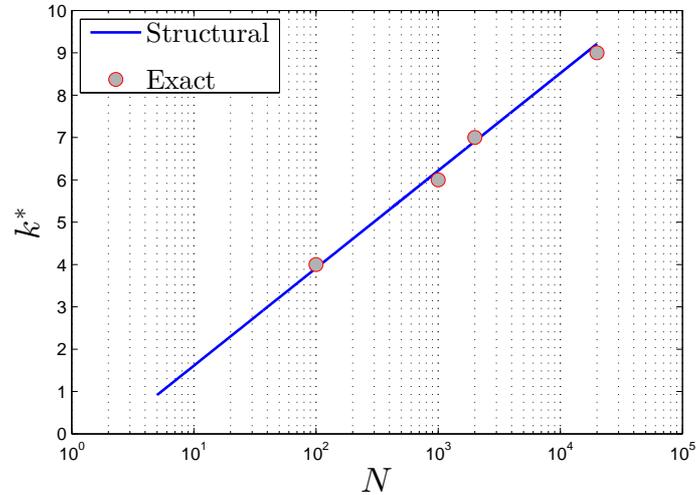}
  \caption{The required number of topological neighbors ($k^*$) in a swarm to reach full
    controllability vs. swarm size ($N$). The blue line corresponds to the
  approximate analytical result from the structural controllability
  analysis. The red dots refer to the result obtained with the exact
  controllability tool.}
\end{figure}

Figure~\ref{figsc} shows the required number $k^*$
of topological agents to achieve full controllability of the swarm based
on the above analytical result. In other words, Fig.~\ref{figsc} provides an
answer to the following question: for a given swarm population
$N$, what is the number of topological neighbors $k^*$ required to confer to
each and every single agent full controllability ``powers'' over all other agents. Moreover,
this approximate analytical result based on the structural controllability is
in very good agreement with those obtained using the exact controllability framework.

\section*{Acknowledgments}
This work was supported by a SUTD-MIT International Design Center (IDC) Grant.

\end{document}